\documentclass[aps,prl,twocolumn,superscriptaddress]{revtex4-1}
\usepackage{graphicx}
\usepackage{dcolumn}
\usepackage{bm}
\usepackage{multirow}
\usepackage{ulem}
\usepackage{color}
\usepackage{subfigure}
\usepackage{amsmath}
\usepackage{upgreek}
\usepackage{amsthm,amsmath,amssymb}

\begin{document}
\title{Multidimensional King-plot analysis for accurate extraction of Cd nuclear charge radii: a challenge for nuclear structure theory}

\author{J. Z. Han}
\affiliation{State Key Laboratory of Precision Measurement Technology and Instruments, Key Laboratory of Photon Measurement and Control Technology of Ministry of Education, Department of Precision Instrument, Tsinghua University, Beijing 100084, China}
\author{C. Pan}
\affiliation{State Key Laboratory of Nuclear Physics and Technology, School of Physics, Peking University, Beijing 100871, China}
\author{K. Y. Zhang}
\affiliation{State Key Laboratory of Nuclear Physics and Technology, School of Physics, Peking University, Beijing 100871, China}
\author{X. F. Yang}
\affiliation{State Key Laboratory of Nuclear Physics and Technology, School of Physics, Peking University, Beijing 100871, China}
\author{S. Q. Zhang}
\affiliation{State Key Laboratory of Nuclear Physics and Technology, School of Physics, Peking University, Beijing 100871, China}
\author{J. C. Berengut}
\affiliation{School of Physics, University of New South Wales, Sydney 2052, Australia}
\author{S. Goriely}
\affiliation{Institut d$'$Astronomie et d$'$Astrophysique, CP-226, Universite Libr\'e de Bruxelles, 1050 Brussels, Belgium}
\author{H. Wang}
\affiliation{School of Economics and Management, Tsinghua University, Beijing 100084, China}
\author{Y. M. Yu}
\email{ymyu@aphy.iphy.ac.cn}
\affiliation{Beijing National Laboratory for Condensed Matter Physics, Institute of Physics, Chinese Academy of Sciences, Beijing 100190, China}
\author{J. Meng}
\email{mengj@pku.edu.cn}
\affiliation{State Key Laboratory of Nuclear Physics and Technology, School of Physics, Peking University, Beijing 100871, China}
\author{J. W. Zhang}
\email{zhangjw@tsinghua.edu.cn}
\affiliation{State Key Laboratory of Precision Measurement Technology and Instruments, Key Laboratory of Photon Measurement and Control Technology of Ministry of Education, Department of Precision Instrument, Tsinghua University, Beijing 100084, China}
\author{L. J. Wang}
\email{lwan@mail.tsinghua.edu.au}
\affiliation{State Key Laboratory of Precision Measurement Technology and Instruments, Key Laboratory of Photon Measurement and Control Technology of Ministry of Education, Department of Precision Instrument, Tsinghua University, Beijing 100084, China}

\begin{abstract}
High-accuracy determination of isotope shift factors, which plays a crucial role in accurate extraction of nuclear charge radius, is well-known to be challenging experimentally and theoretically.
Nonetheless, based on an accurate measurement of the isotope shifts for the Cd$^+$ $5s~^2S_{1/2}-5p~^2P_{3/2}$ and the Cd $5s^2~^1S_0-5s5p~^1P_1$ transition, a multidimensional King-plot analysis is performed to extract the atomic field shift and mass shift factors accurately. The results are further cross-checked against results from atomic structure calculations using a high-accuracy configuration interaction and many-body perturbation theory. Combined with previous isotope shifts, nuclear charge radii for $^{100-130}$Cd of the highest accuracy are reported. For the neutron-rich region, accuracies for the charge radii are improved by nearly one order of magnitude. This work provides a coherent and systematic extraction of Cd nuclear charge radii from isotope shifts. The results offer stringent testing and possible challenges to the latest advances in nuclear theory and excellent benchmarking to the atomic structure calculations.
\end{abstract}
\date{\today}

\maketitle
\textit{Introduction.—}
The nuclear charge radius is a fundamental property of an atomic nucleus. It serves as an important probe in various nuclear phenomena, such as the occurrence of magic numbers \cite{koszorus2021charge, gorges2019laser, goodacre2021laser}, shape staggering \cite{marsh2018characterization} and evolution \cite{de2007nuclear, yordanov2012nuclear}, shape coexistence \cite{yang2016isomer}, proton \cite{geithner2008masses} and neutron \cite{sanchez2006nuclear, noertershaeuser2009nuclear} halos, and neutron skins \cite{hagen2016neutron}. Moreover, the subtle evolution of charge radii along isotopic chains provides a stringent test and challenge for nuclear models. Therefore, for both experimental and theoretical nuclear physics, a highly accurate extraction of nuclear charge radii is paramount.
With the development of laser spectroscopy technology, the relative changes in the nuclear charge radii may be precisely determined from isotope shifts (ISs) \cite{gebert2015precision, hammen2018calcium, koszorus2021charge}. Most radio-nuclide measurements currently rely on laser spectroscopy. However, the extraction of the nuclear charge radius relies on atomic field shift factors $F$ and mass shift factors $K$. At present, the accuracy of atomic factors has become the most critical limiting factor in extracting the nuclear charge radius and highlights the importance of determining them with high accuracy and reliability. 

Cd nuclei have attracted wide attention because of their special attributes: i) Cd nuclei contain 48 protons, two proton holes below the $Z=50$ proton magic number---nuclei near the shell having richer properties; ii) a radioactive ion beam facility can provide Cd isotopes with mass numbers ranging from 100 to 130 \cite{hammen2018calcium}--- many isotopes of an atom provide considerable sample information regarding the changes in nuclear properties with neutron number; iii) in atomic physics, Cd/Cd$^+$ systems have many spectral transitions suitable for optical measurements taken using a Cd$^+$ microwave clock \cite{zhang2012high, wang2013high, miao2015high} and sympathetic cooling technology \cite{zuo2019direct, han2021toward}---here, the measurement accuracy of the Cd$^+$ ISs may be further improved. In the past, many groups have measured the optical ISs of Cd isotopes and extracted nuclear charge radii \cite{gillespie1975measurements, wenz1981subshell, buchinger1987n, angeli2013table, fricke2004nuclear, yordanov2013spins, hammen2018calcium}. However, the results extracted from independent groups are not consistent (e.g. Refs.~\cite{fricke2004nuclear, hammen2018calcium, angeli2013table}). For Cd nuclei, with nucleon numbers near the magic numbers, $50$ and $82$, the accuracy of the nuclear charge radii is relatively low \cite{hammen2018calcium}.

Here, for $^{100-130}$Cd nuclear charge radii, we present our latest results which currently have the highest accuracy. The atomic $F$ and $K$ factors used to extract the nuclear radii are obtained experimentally from a multidimensional King-plot analysis (MKA) and were cross-checked using theoretical atomic structure calculations. In experiments, laser-induced fluorescence (LIF) from sympathetically-cooled large Cd$^+$ ion crystal and saturated absorption spectroscopy are used to measure the ISs in the Cd$^+$ $5s~^2S_{1/2}\rightarrow5p~^2P_{3/2}$ transition line (214.5 nm) and the Cd $5s^2~^1S_0\rightarrow 5s5p~^1P_1$ transition line (228.8 nm). Combined with the IS results of the Cd $5s5p~^3P_2 - 5s6s~^3S_1$ transition line (508.6 nm) and the nuclear size parameter $\lambda$, MKA plots were used to extract the atomic factors. From theory, high-accuracy atomic structure calculations were performed using the configuration interaction and many-body perturbation theory (CI+MBPT) approach to obtain the atomic factors for the typical Cd/Cd$^+$ transitions. These atomic factor results ensure the accuracy and reliability of our latest calculations of the $^{100-130}$Cd nuclear charge radii, especially for those nuclei with neutron numbers approaching magic numbers, $N=50$ and $82$. This Letter reports on a more consistent way to determine the atomic factors and extract the nuclear charge radii through IS calculations. High-accuracy data of nuclear charge radii will prompt further testing and provide guidance of nuclear theory.

\begin{figure}
\centering
\resizebox{0.45\textwidth}{!}{
\includegraphics{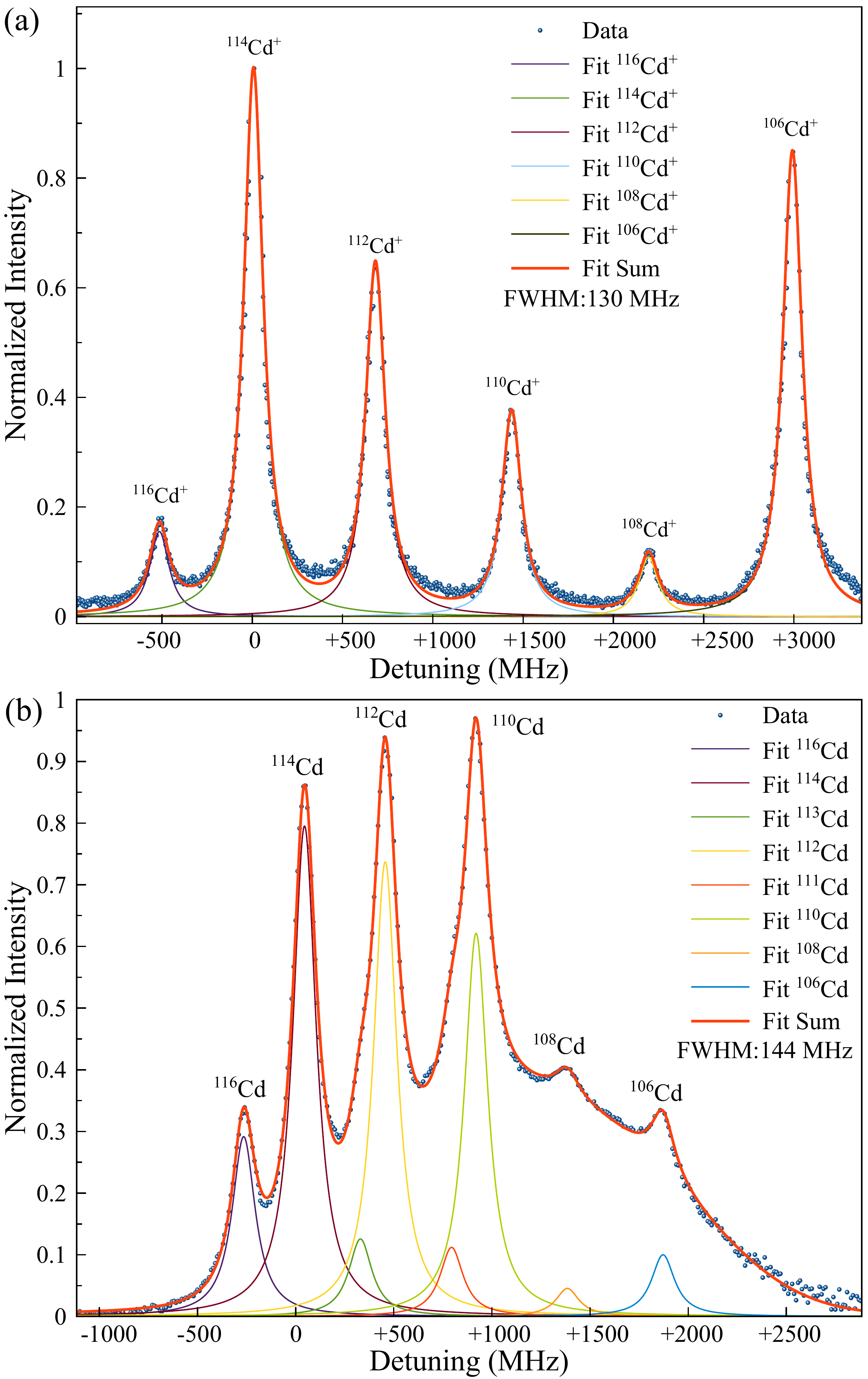}
}
\caption{Measurement results for (a) the Cd$^+$ $5s~^2S_{1/2}\rightarrow5p~^2P_{3/2}$ transition line (214.5 nm) fitted using Voigt profiles, and (b) the Cd $5s^2~^1S_0\rightarrow 5s5p~^1P_1$ transition line (228.8 nm) fitted using the Lorentz profile (signal) and the Doppler profile (background, not shown). For both plots, each measurement point is the average of three measurements, and each colored line represents the fitted curve of an isotope, and the red solid line represents the total fitted curve.}
\label{fig:Cd+}
\end{figure}

\textit{Experiment.—} 
The MKA starts from high-accuracy IS measurements of the transitions for two Cd systems specifically, the Cd$^+$ 214.5 nm and Cd 228.8 nm transition lines. The ISs of the Cd$^+$ 214.5 nm line were measured using LIF spectroscopy. In a linear Paul trap, we prepared two-species crystals consisting of $2\times10^5$ Ca$^+$ and $4\times10^5$ Cd$^+$ ions (see Ref. \cite{han2021toward} for details). The coolant ions, Ca$^+$ ions, were Doppler-cooled by lasers, and the Cd$^+$ ions were sympathetically cooled to less than 0.5 K through the Coulomb interaction between ions. The ISs of the transition line for cooled Cd$^+$ were measured by scanning the frequencies in the weak 214.5-nm probe laser beam. The frequencies of the probe laser beam spans 5 GHz thus permitting the measurement of all frequency resonances of the Cd$^+$ isotopes of even-numbered neutrons. The frequencies of the laser beam were measured using a high-precision wavemeter, which was calibrated against an optical frequency comb referenced to a hydrogen maser. The measured LIF spectrum is shown in Fig. \ref{fig:Cd+}(a). The ISs of the Cd 228.8 nm line were measured through SAS using a hollow cathode lamp. The 228.8-nm laser beam was split into two beams, a pump and a probe, which were subsequently configured to overlap at the center of the lamp. From the measured SAS data taken with a sample with a natural abundance of Cd [Fig. \ref{fig:Cd+}(b)], the source of uncertainty of the IS results is mainly statistical uncertainty. With many of the systematic effects being common to all the neutron-even isotopes and the potential systematic shifts obtained from differences of line frequencies, many potential systematic shifts are common and cancel to a high degree. The measurement results of the ISs of Cd$^+$ 214.5 nm line and Cd 228.8 nm line are listed in Table \ref{tab:IS}. For the former, our results are consistent with those of Hammen and colleagues \cite{hammen2018calcium}, although the accuracy is improved by 2--6 times. For the Cd 228.8 nm line, our results are to our best knowledge the first high-precision IS measurements, being 30 times more accurate than earlier experimental results \cite{kelly1961isotope}.

\begin{table}
\caption{Measured values of the isotope shifts of the Cd$^+$ $5s~^2S_{1/2}-5p~^2P_{3/2}$ (214.5 nm) and Cd $5s^2~^1S_0- 5s5p~^1P_1$ (228.8 nm) transition lines compared with Ref. \cite{hammen2018calcium} and \cite{kelly1961isotope}. } \label{tab:IS}
{\setlength{\tabcolsep}{6pt}
\begin{tabular}{llllllll}\hline\hline
$A'$ &\multicolumn{2}{c}{$\delta\nu^{114,A'}_{214.5}$ (MHz)} &\multicolumn{2}{c}{$\delta\nu^{114,A'}_{228.8}$ (MHz)} \\ 
&This work & \cite{hammen2018calcium} & This work & \cite{kelly1961isotope} \\\hline
116 &-521.2(1.2) &-526.5(2.7) & -310.1(1.2) & \\ 
112 &~675.4(1.2) &~674.6(2.7) &~411.2(1.2) & 396(30)\\ 
110 &1429.8(1.2) &1432.2(3.8) & ~874.2(1.2) & 906(35) \\
108 &2187.8(1.2) &2194.0(5.0) & 1339.0(2.1) & \\ 
106 &2984.9(1.2) &2991.1(6.6) &1826.6(1.2) \\\hline\hline
\end{tabular}}
\end{table}

\textit{Multidimensional King-plot Analysis.—} 
According to IS theory \cite{breit1958theory, king2013isotope}, the relationship between isotope shift $\delta \nu^{A,A'} = \nu^A - \nu^{A'}$ and nuclear size $\lambda$ is
\begin{equation}
\mu \delta \nu^{A,A'}=F\cdot \mu \lambda^{A,A'}+K,
\label{Eq:IS}
\end{equation}
where $\mu=m_A\cdot m_{A'}/(m_{A}-m_{A'})$. The nuclear size data are obtained from muonic atom spectroscopy and electron scattering experiments. The KA procedure introduces the $F$ and $K$ factors determined from the slope and intercept of the fitted curve of the $\mu \delta \nu^{A,A'}$ and $\mu \lambda^{A,A'}$ data. This procedure mitigates inaccuracies of the atomic model calculation. However, it strongly relies on the precision of $\lambda^{A,A'}$ obtained through muonic atom spectroscopy and electron scattering experiments. Previously, based on the ISs of the Cd$^+$ 214.5 nm line and the $\lambda$ input, Hammen and colleagues performed a two-dimensional King-plot analysis (2D KA) and obtained values of $K$ and $F$ for the Cd$^+$ 214.5-nm transition line, but with large uncertainties \cite{hammen2018calcium}. These uncertainties have become the main source of uncertainty in Cd nuclear charge radii. In the MKA method, we used the ISs of the Cd$^+$ 214.5-nm line, of the Cd 228.8-nm line, of Cd 508.6-nm line \cite{hammen2018calcium}, and values of the nuclear size $\lambda$ \cite{fricke2004nuclear} as input to reduce the uncertainties of the extracted $K$ and $F$ factors. In the MKA method, the $K$ and $F$ factors are constrained by the assumption underlying Eq. (\ref{Eq:IS}). There are four experimental data points corresponding to isotope pairs $(A,A')=(110,108),(112,110), (114,112), (116,114)$, which should lie on a straight line in the four-dimensional isotope space formed from $\mu\delta\nu^{A,A'}_{214.5}$, $\mu\delta\nu^{A,A'}_{228.8}$, $\mu\delta\nu^{A,A'}_{508.6}$, and $\mu\lambda^{A,A'}$. The objective of the MKA method is to ensure all correlations and uncertainties are made correctly thereby enabling the $K$ and $F$ factors to be extracted more accurately and reliably.

The procedures of the MKA method are as follows:
i) The uncertainties of the experimental data points $D_1,\cdots,D_k$ in the multidimensional space are assumed to be independent of each other and comply with the multivariate normal distribution. The covariance matrix of the multivariate normal distribution are the corresponding measurement uncertainties \cite{gebert2015precision};
ii) Two arbitrary Monte Carlo points are generated within the 1$\sigma$ range of two selected experimental points (e.g. $D_1$ and $D_k$), and a multidimensional straight line $l$ is generated using these two Monte Carlo points;
iii) Projection points (labeled $P_1,\cdots,P_k$) for each data point ($D_1,\cdots,D_k $) on the straight line $l$ are generated. The probability density of each data point at the corresponding projection points $f_{D_1}(P_1),\cdots,f_{D_k}(P_k)$ is obtained;
iv) The product of the probability densities of each projection point, $f=f_{D_1}(P_1)\cdot f_{D_2}(P_2)\cdot\cdots\cdot f_{D_k}(P_k)$, is used as the selection criterion. We generate $N$ sets of Monte Carlo samples (in this work, we chose $N=3\times10^7$) and selected $M$ sets of samples that meet the condition $f/f_{max}>1\sigma$, where $f_{max}$ is the maximum probability density of all $N$ sets of Monte Carlo samples;
v) The selected samples generate $M$ sets of multidimensional straight lines $l$ that were then used to extract the $K$ and $F$ factors. These factors are thus obtained from the probability density of the data points and have uncertainties with 1$\sigma$ range.

The MKA provides model-independent atomic factors for the Cd$^+$ 214.5-nm, Cd 228.8-nm and Cd 508.6-nm transition lines (see Table \ref{tab:FSMS}). Comparing the results of the MKA and 2D KA, and with the addition of two other dimensions, the accuracy of the $K$ and $F$ factors was improved by about 3.5 times, thereby confirming that the MKA method functions as claimed. The MKA method is expandible to higher dimensions and may be applied to elements with at least three measured nuclear size parameters $\lambda$. More high-accuracy IS measurements of other transition lines in the Cd ions may further improve the accuracy of both $K$ and $F$ in the MKA. Comparing the 2D KA results with those in Ref. \cite{hammen2018calcium}, we find that using the more accurate IS results of the Cd$^+$ 214.5-nm transition line, the extracted atomic factors become twice as accurate, highlighting the importance of further improving the accuracy of the IS measurements for stable nuclei. Comparing the MKA results with those of Hammen and colleagues, the accuracy of the atomic factors is improved six-fold, ensuring that we can extract Cd nuclear charge radii of greater accuracy. The MKA results also provide excellent reference data for benchmarking and cross-checking atomic structure calculations.

\begin{table}
\caption{Field shift ($F$) and mass shift ($K$) factors for Cd$^+$ and Cd obtained by the MKA method, 2D KA method, and the CI+MBPT calculations. The atomic factors for several typical transition lines of the Cd/Cd$^+$ systems obtained from CI+MBPT calculations are also listed.} \label{tab:FSMS}
{\setlength{\tabcolsep}{4pt}
\begin{tabular}{lll}\hline\hline
$F$ (MHz/fm$^2$) &$K$ (GHz u) &Source \\\hline 
\multicolumn{3}{l} {Cd$^+$ $5s~^2S_{1/2}-5p~^2P_{3/2}$ (214.5 nm) } \\

-6192(320) &1851(315) &MKA (This work) \\
-6176(1107) & 1831(1093)&2D KA (This work) \\
-6144(300) &1667(300) &CI+MBPT (This work)\\ 
-6260(1860) &1860(1920) &2D KA \cite{hammen2018calcium} \\

\multicolumn{3}{l} {Cd $5s^2~^1S_{0}-5s5p~^1P_{1}$ (228.8 nm) } \\
-4140(214) &1491(211) &MKA (This work)\\
-4129(741)& 1477(731)&2D KA (This work) \\
-4024(200) &1428(220) &CI+MBPT (This work)\\

\multicolumn{3}{l} {Cd $5s5p~^3P_{2}-5s6s~^3S_{1}$ (508.6 nm) } \\
1194(64) &-20(63) &MKA (This work)\\
1216(226)& -38(222)&2D KA (This work) \\
1228(60) &-63(400) &CI+MBPT(This work) \\

\multicolumn{3}{l} {Cd$^+$ $5s~^2S_{1/2}-5p~^2P_{1/2}$ (226.5 nm) } \\
-6067(300) &1770(300) &CI+MBPT (This work)\\ 
-6077(734) &1468(734) &semi-empirical \cite{bishop1971isotope} \\ 
-6174(778) &558(385) &semi-empirical \cite{bauche1985analysis} \\ 

\multicolumn{3}{l} {Cd $5s^2~^1S_{0}-5s5p~^3P_{1}$ (326.1 nm) } \\
-4559(230) &1865(400) &CI+MBPT (This work)\\ 
-4420 &1717(330) &2D KA \cite{fricke2004nuclear} \\ 
-3910(460) &876(230) &semi-empirical \cite{angeli2013table} \\ 
-3900(460) &809(407) &semi-empirical \cite{buchinger1987n} \\

\hline\hline
\end{tabular}}
\end{table}

\textit{Atomic structure calculation.—}
The calculations of the $F$ and $K$ factors set strong challenges for many-body atomic theory. The mass shift, in particular, has proven very difficult to calculate even for single-valence-electron Cd$^+$ isotopes. We used the CI+MBPT method, implemented in the AMBiT software \cite{kahl2019ambit}, to calculate the atomic factors. The CI+MBPT calculations start from a self-consistent Dirac--Fock--Breit (DFB) calculation employing the potential of a closed-shell Pd-like core. Next, a large valence basis of one-particle orbitals is generated by diagonalizing a set of B-splines over the one-electron DFB operator. Then, a set of many-electron configurations for the CI expansion are constructed with these orbitals. For neutral Cd, the CI configurations include all valence excitations up to level $20spdf$. For Cd$^+$, we also need to access the hole states, so we employ a particle--hole CI+MBPT~\cite{berengut16pra}, for which the core $4d$ shell is unfrozen and included in the CI configurations. The configuration set is generated from all single and double excitations up to $10spdf$ from leading configurations $5s$, $5p$, $5d$, and $4d^{-1}\,5s^2$. Correlations with core shells that are not included in the CI expansion are accounted for using the MBPT for both Cd and Cd$^+$. This includes excitations to virtual orbitals up to $30spdfg$.

The finite-field approach is implemented in the $F$ and $K$ factors calculation of the AMBiT software \cite{berengut2003isotope}. Therein, the operators for $F$ and $K$ are added to the Coulomb potential from the very beginning of the DFB calculation. An IS scaling factor $\Gamma$ ahead of the $F$ and $K$ operators is varied, and the level energies $E$ are calculated as a function of $\Gamma$. The gradient of an $E$ versus $\Gamma$ graph then gives us the field and mass shift factors. Values of the $F$ and $K$ factors for Cd$^+$ and Cd obtained by the CI+MBPT calculations are listed in Table \ref{tab:FSMS}. For the Cd$^+$ 214.5-nm, Cd 228.8-nm, and Cd 508.6-nm transition lines, the CI+MBPT results show excellent consistency with the MKA results, thus highlighting the high accuracy and reliability of this method. The CI+MBPT method has also checked for accuracy and reliability in calculations of the atomic factors of other neutral atoms and monovalent ions systems (see for example Ca$^+$ \cite{gebert2015precision, solaro2020improved, muller2020collinear} and Yb$^+$ \cite{counts2020evidence}). Cross-checking of the atomic factors obtained using the CI+MBPT and MKA further confirms the reliability of our results.

\begin{table}
\caption{Extracted changes in the mean-square charge radii $\delta \langle r^2\rangle $ (fm$^2$) relative to $^{114}$Cd and the nuclear charge radius $R_{ch}$ (fm) of $^{100-130}$Cd. We used the Cd$^+$ 214.5-nm IS data of this work for those isotopes tagged by a superscripted $*$. For those isotopes tagged by a $\dagger$ and those untagged isotopes, we used respectively the Cd 508.6-nm IS data and the Cd$^+$ 214.5-nm IS data of Hammen and colleagues \cite{hammen2018calcium}.
The various uncertainties associated with the $\delta \langle r^2\rangle$ and $R_{ch}$ values are given in parentheses: the first relate to uncertainties arising from the IS measurements, the second relate to systematic errors stemming from the calculation of the atomic $F$ and $M$ factors, and the third for the $R_{ch}$ values are the uncertainties originating from the $^{114}$Cd measurements. The isotope-dependent linear transformation parameters $\alpha$ (fm$^2$ u) and the corresponding isotope-dependent mass shift factors $K_{\alpha}$ (GHz u) are also listed.
}
\label{tab:rms charge radii}
{\setlength{\tabcolsep}{2pt}
\begin{tabular}{lllllllllll}\hline\hline
N & $\alpha$ & $K_{\alpha}$ & $\delta \langle r^2\rangle$ & $R_{ch}$ \\\hline
52 & 1061 & -4719(34) & -1.4350(20)(85) & 4.45372(22)(96)(100) \\
53 & 1061 & -4719(34) & -1.3196(18)(78) & 4.46665(20)(88)(100) \\
54 & 1036 & -4564(29) & -1.1533(16)(57) & 4.48522(18)(63)(100) \\
55 & 1040 & -4589(30) & -1.0551(15)(54) & 4.49616(17)(60)(100) \\
56 & 1017 & -4447(26) & -0.9105(14)(39) & 4.51221(15)(43)(100) \\
57 & 1015 & -4434(26) & -0.8302(12)(40) & 4.52110(13)(44)(100) \\
58$^*$ & 1006 & -4378(25) & -0.6991(2)(28) & 4.53558(3)(31)(100) \\
59 & 1026 & -4502(27) & -0.6298(10)(29) & 4.54321(11)(32)(100) \\
60$^*$ & 1004 & -4366(24) & -0.5131(2)(20) & 4.55604(3)(22)(100) \\
61 & 1034 & -4552(28) & -0.4488(7)(22) & 4.56308(8)(24)(100) \\
62$^*$ & 1003 & -4360(24) & -0.3355(2)(13) & 4.57548(3)(14)(100) \\
63 & 1089 & -4892(41) & -0.2911(5)(21) & 4.58033(6)(23)(100) \\
64$^*$ & 990 & -4279(24) & -0.1603(2)(6) & 4.59459(3)(7)(100) \\
65 & 1219 & -5697(79) & -0.1160(4)(14) & 4.59941(4)(15)(100) \\
66 & \multicolumn{2}{c}{Reference} & 0(0)(0) & 4.61200(0)(0)(100) \\
67 & 758 & -2843(76) & 0.0418(5)(13) & 4.61653(5)(14)(100) \\
68$^*$ & 917 & -3827(32) & 0.1331(2)(9) & 4.62640(3)(10)(100) \\
69$^{\dagger}$ & 855 & 1000(9) & 0.1700(20)(23) & 4.63040(22)(25)(100) \\
70$^{\dagger}$ & 884 & 1035(8) & 0.2424(21)(25) & 4.63821(23)(27)(100) \\
71$^{\dagger}$ & 858 & 1004(9) & 0.2811(29)(38) & 4.64238(31)(41)(100) \\
72$^{\dagger}$ & 865 & 1012(9) & 0.3405(20)(43) & 4.64876(22)(46)(100) \\
73 & 847 & -3394(50) & 0.3737(11)(55) & 4.65233(11)(59)(100) \\
74$^{\dagger}$ & 849 & 993(10) & 0.4276(25)(63) & 4.65813(27)(67)(100) \\
75 & 834 & -3313(53) & 0.4553(12)(76) & 4.66110(13)(81)(100) \\
76 & 838 & -3338(52) & 0.5081(12)(81) & 4.66676(13)(87)(100) \\
77 & 822 & -3239(57) & 0.5295(14)(97) & 4.66906(15)(104)(100) \\
78 & 828 & -3276(55) & 0.5823(15)(102) & 4.67471(16)(109)(100) \\
79 & 811 & -3171(60) & 0.5941(16)(121) & 4.67597(18)(129)(100) \\
80 & 821 & -3233(57) & 0.6561(17)(122) & 4.68259(18)(130)(100) \\
81 & 790 & -3041(66) & 0.6316(20)(153) & 4.67997(21)(163)(100) \\
82 & 804 & -3128(62) & 0.6992(20)(151) & 4.68719(21)(161)(100) \\
\hline\hline
\end{tabular}}
\end{table}

\textit{Nuclear Charge Radii of $^{110-130}$Cd.—} 
Using the MKA method and our measured IS of the Cd$^+$ 214.5-nm and Cd 228.8-nm transition lines, the $^{100-130}$Cd nuclear charge radii can be extracted with greater accuracy. An isotope-dependent linear transformation $\mu\lambda_{\rm{Cd}}\rightarrow (\mu\lambda_{\rm{Cd}}-\alpha)$ is used to reduce the extracted uncertainties \cite{hammen2018calcium}. For each isotope, the nuclear size factor $\mu\lambda_{\rm{Cd}}$ is translated to the position at which the slope and intercept of the fitted curve have the least correlation. Then, the MKA method is used to extract the mass shift factor $K_{\alpha}$ from the set of data points ($\mu\delta\nu_{214.5},~\mu\delta\nu_{228.8},~\mu\delta\nu_{508.6}, ~\mu\lambda_{\rm{Cd}}-\alpha$). The changes in the nuclear charge radii $\delta\langle r^2 \rangle$ are subsequently extracted using
\begin{eqnarray}
\lambda^{A,A'}&=&\frac{\delta \nu_{IS}^{A,A^{'}}
-K_{\alpha}/\mu^{A,A^{'}}}{F}+\frac{\alpha}{\mu^{A,A^{'}}}\nonumber\\
&=&\delta\langle r^2\rangle^{A,A^{'}}+\frac{C_2}{C_1}\delta\langle r^4 \rangle^{A,A^{'}}+\frac{C_3}{C_1}\delta\langle r^6 \rangle^{A,A^{'}},
\end{eqnarray}
where $C_{1,2,3}$ denote the coefficients associated with the contributions of the IS from higher radial moments \cite{seltzer1969k}. For the radioactive isotopes, a contribution of $-2.7\%$ is assumed \cite{fricke2004nuclear}.

\begin{figure}
\centering
\resizebox{0.4\textwidth}{!}{
\includegraphics{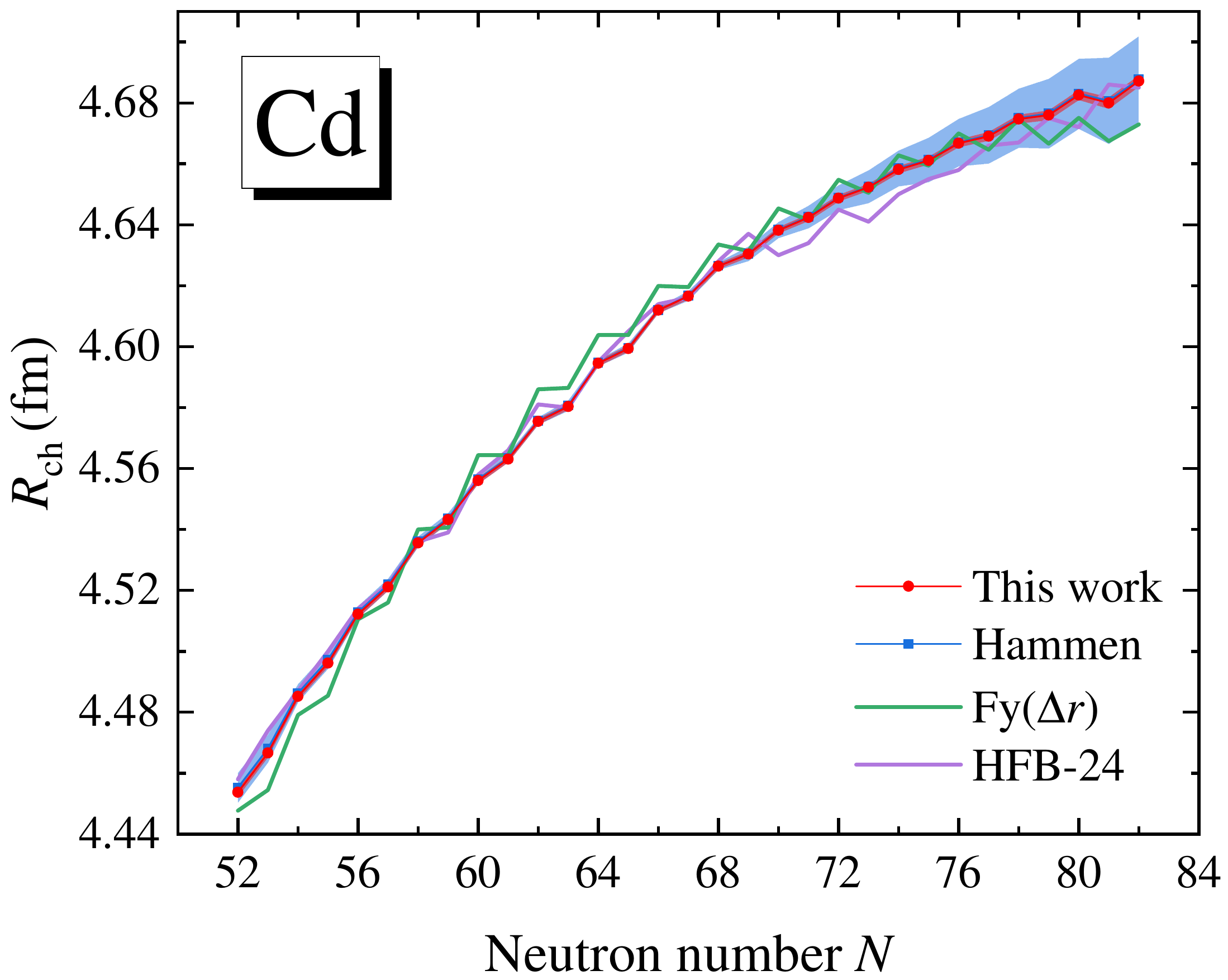}
}
\caption{Comparison of the accuracy for the $^{100-130}$Cd nuclear charge radii of this work and the results in Ref.~\cite{hammen2018calcium}. Theoretical predictions obtained using Fayans functionals Fy($\Delta r$) and the Skyrme--Hartree--Fock--Bogoliubov model HFB-24 are shown for comparison.}
\label{Fig:Rc}
\end{figure}

For the naturally stable isotopes with $A=$110, 112, 114, 116, and 118, the new measurements of the IS for the Cd$^+$ 214.5-nm transition line have improved accuracies; for the other isotopes of Cd, the measurements by Hammen and colleagues give IS values of sufficient precision \cite{hammen2018calcium}. These results lead to a new determination of the $^{100-130}$Cd nuclear charge radii based on our $F$ and $K$ values obtained by the MKA method. The recommended values of $\delta\langle r^2 \rangle$ and nuclear charge radii $R_{ch}$ are listed in Table \ref{tab:rms charge radii}. The isotope-dependent linear transformation used in extracting the nuclear charge radius from the IS data eliminates the quadratic trend in the uncertainty, and the more accurate values of $F$ and $K$ yield more precise determination of $R_{ch}$, especially for isotopes from the neutron-deficient ($N<60$) and neutron-rich ($N>70$) regions (Fig. \ref{Fig:Rc}). The results reported in this Letter agree with those of previous work lying well within the errors of the latter and with much improved accuracy. For the neutron-rich region, the charge-radii accuracy has been improved by nearly one order of magnitude. The improved accuracy highlights the odd-even staggering (OES) of the charge radii in Cd isotopes for $N>70$, the phase of which was uncertain in other work.

The present high-precision results of the Cd charge radii set a formidable benchmark for the latest advances in nuclear theory. In Fig. \ref{Fig:Rc}, we include two modern theoretical predictions for comparison, the Fayans functional Fy($\Delta r$) \cite{reinhard2017toward, hammen2018calcium} and one of best fits of the Skyrme--Hartree--Fock--Bogoliubov mass formulas, HFB-24 \cite{goriely2013further}. Note that other predictions from different density functional calculations have been compared in Ref. \cite{hammen2018calcium}, but most fail to reproduce the isotopic trend as a whole and the OES in detail. The Fayans functional, which includes gradient terms in the surface term and the pairing functional \cite{fayans2000nuclear, reinhard2017toward} reproduce satisfyingly the experimental data. For this parametrization, which is tuned with the Ca chain, the functional performs well for the Cd chain, capturing not only the isotopic trend but also the correct phase of the OES \cite{hammen2018calcium}.

Nevertheless, the Fy($\Delta r$) results for the neutron-rich Cd nuclei near the magic number $N=82$, which were within the errors of previous experimental data, are now beyond the high-accuracy results obtained in this work. As for the K isotopes, the amplitudes of the OES in general are also overestimated by Fy($\Delta r$), \cite{koszorus2021charge}. The HFB-24, in which the strength in contact pairing for neutrons (protons) depends on the neutron and proton densities, provides a remarkable quantitative description of the experimental data. Nevertheless, some apparent discrepancies exist in the OES, e.g., the opposite phase near $N=80$. Other best Skyrme HFB-n mass formulas, such as HFB-25, HFB-29, and HFB-31 \cite{goriely2013further, goriely2015further, goriely2016further}, show similar performances as HFB-24 and are thus not shown explicitly in Fig. \ref{Fig:Rc}. Therefore, the present Cd charge radii with its improved accuracy, particularly for the neutron-rich region, are set to provide stringent tests and challenges for advanced nuclear models.

\textit{Conclusion.—}
High accuracy measurements of the IS of the Cd$^+$ $5s~^2S_{1/2}-5p~^2P_{3/2}$ (214.5 nm) and the Cd $5s^2~^1S_0-5s5p~^1P_1$ (228.8 nm) transition lines were reported. Combined with the nuclear size parameter $\lambda$ and the ISs of the Cd $5s5p~^3P_2 - 5s6s~^3S_1$ (508.6 nm) transition line, a MKA was performed to extract the $F$ and $K$ factors of these transition lines. The MKA method placed strong constraints on linear fits, and the uncertainties for the $F$ and $K$ factors are well-determined by the multivariate normal distribution. For the Cd$^+$ $5s~^2S_{1/2}-5p~^2P_{3/2}$ transition line, the accuracy of the extracted $F$ and $K$ factors from the MKA were improved six-fold compared with previous results and furthermore were cross-checked in high-accuracy CI+MBPT calculations. From the improved values for $F$ and $K$, more accurate values of the nuclear charge radii of $^{100-130}$Cd were extracted. The results are consistent with previous work, being well within previous errors but with a greater accuracy, especially for isotopes of the neutron-rich and neutron-deficient regions. The high-precision nuclear charge radii of $^{100-130}$Cd reported in this Letter set stringent constraints and pose challenges to current nuclear theory, in which state-of-the-art density functional calculations, HFB-24 and Fy($\Delta r$), fail to reproduce perfectly currently-accepted results.
In addition, accurate nuclear charge radii may help in future investigations of Cd shell closure when extending the neutron number up to magic number $N=82$ if new neutron-rich Cd isotopes are fabricated.

\textit{Acknowledgments.—}
We thank V. A. Dzuba, W. N\"ortersh\"auser, D. T. Yordanov, P. O. Schmidt and T. Li for helpful discussions. This work is supported by the National Key R\&D Program of China (Nos. 2016YFA0302100, 2017YFE0116700, 2018YFA0404400), National Natural Science Foundation of China (Nos. 12073015, 11935003, 11875075) and Beijing Natural Science Foundation (1202011).

\bibliography{apssamp}
\end{document}